\documentclass[11pt,dvips]{article}
\usepackage{graphicx}
\usepackage{amsmath}
\usepackage{amssymb}
\usepackage{latexsym}
\usepackage{color}
\begin{document}
\thispagestyle{empty}
\begin{center}

\vspace{1.8cm}

 \Large{\bf Global quantum correlations in tripartite nonorthogonal  states and monogamy properties }\\

\vspace{1.5cm}

{\bf M. Daoud}$^{a,b}$ , {\bf R. Ahl Laamara}$^{c,d}$ , {\bf R.
Essaber} $^c$  and {\bf W. Kaydi}$^{c}$

\vspace{0.5cm}
$^{a}${\it Department of Physics, Faculty of Sciences, University Ibnou Zohr,\\
 Agadir,
Morocco}\\[1em]
$^{b}${\it Abdus Salam International Centre for Theoretical Physics,\\
Trieste,
Italy}\\[1em]

$^{c}${\it LPHE-Modeling and Simulation, Faculty of Sciences,
University
Mohamed V,\\ Rabat, Morocco}\\[1em]

$^{d}${\it Centre of Physics and Mathematics,
CPM, CNESTEN,\\ Rabat, Morocco}\\[1em]

\vspace{1.5cm} {\bf Abstract}
\end{center}
\baselineskip=18pt
\medskip

A global measure of quantum correlations for  tripartite
nonorthogonal states is presented. It is introduced as the  overall
average  of the pairwise correlations existing in all possible
partitions. The explicit expressions for the global measure are
derived for squared concurrence, entanglement of formation, quantum
discord and its geometric variant. As illustration, we consider even
and odd three-mode Schr\"odinger cat states  based on Glauber
coherent states. We also discuss limitations to sharing quantum
correlations known as monogamy relations.

\newpage
\section{Introduction and motivations}

Remarkable achievements in characterizing, identifying and
quantifying quantum correlations in bi-partite quantum systems were
accomplished in the last two decades
\cite{NC-QIQC-2000,Alber-QI2001,Vedral-RMP-2002,Horodecki-RMP-2009,Guhne}
(for a recent review see \cite{vedral-modi}).
Quantum entanglement is an useful resource for quantum information
processing such as quantum teleportation \cite{Ben1}, superdense
coding \cite{Ben2}, quantum key distribution \cite{Eckert},
telecloning \cite{Murao} and many more. Until some time ago,
entanglement was usually regarded as synonymous of quantum
correlation and subsequently considered as the only type of
nonclassical existing in a multipartite quantum system. However,
quantum entanglement does not account for all nonclassical aspects
of quantum correlations and unentangled mixed states can possess
quantum correlations. In this respect,  other measures of quantum
correlations beyond entanglement were studied. The most popular
among them is quantum discord introduced in
\cite{Ollivier-PRL88-2001,Vedral-et-al}.
 It coincides with entanglement of formation for pure
 states. For mixed states, the explicit evaluation of quantum discord involves
  potentially complex optimization procedure which was achieved for
  a limited set of two qubit systems \cite{Luo,Ali,Shi1,Girolami,Shi2,Rachid1,Rachid2}. To overcome this problem
  an alternative geometrized variant of quantum discord was introduced \cite{Dakic2010}. Nowadays, entanglement of formation
\cite{Wootters98}, quantum discord
\cite{Ollivier-PRL88-2001,Vedral-et-al} and its geometric variant
\cite{Dakic2010} are typical examples of bipartite measures commonly
used to decide about the presence of quantum correlations in a
bipartite quantum system. \\

In other hand, the characterization of genuine correlations in
multipartite quantum systems encounters many conceptual obstacles
and the extension of usual bipartite measures for many-particles
systems is not well understood \cite{vedral-modi}. Despite many
efforts regarding this problem
\cite{Zhou,Kaszlikowski,Bennett,Giorgi1,Li-Luo}, there are still
many unsolved issues. The main motivation behind these efforts
relies upon the recent experimental results reporting the creation
and manipulation of macroscopic quantum states and highly correlated
atomic ensembles such as spin squeezed states
\cite{Hald,Kuzmich,Meyer}. Accordingly, different approaches to
quantify multipartite correlations in quantum systems have been
proposed in the litterature \cite{Chakrabarty,Rulli,Z-H Ma}. In
particular, Rulli and Sarandy \cite{Rulli} defined the multipartite
measure of quantum correlation as the maximum of the quantum
correlation existing between all possible bipartition of the
multipartite quantum system. In this paper, paralleling the
treatment discussed in \cite{Z-H Ma}, we define the global quantum
correlation present in a tripartite system $ABC$ of type
(\ref{eq:main}) as the sum of the correlations of all possible
bi-partitions. Explicitly, it is given by
\begin{eqnarray}\label{Qtotal}
Q_{(A,B,C)} &=& \frac{1}{12} \bigg( Q_{AB} +  Q_{BA} + Q_{AC} +
Q_{CA} + Q_{BC} + Q_{CB}  \nonumber \\ &+& Q_{A( BC)} + Q_{(BC)A} +
Q_{B( AC)} + Q_{(AC)B} + Q_{C( AB)} + Q_{(AB)C}\bigg)
\end{eqnarray}
where the measure $Q$ stands for concurrence, entanglement of
formation, entropy based quantum discord or geometric quantum
discord.\\

Another important feature appearing in investigating  multipartite
quantum correlations is the so-called monogamy relation which
imposes severe restriction of shareability of quantum correlations
in a quantum system comprising three or more parts. The monogamy
relation was first considered by Coffman, Kundo and Wootters in 2001
\cite{Coffman} in analyzing the distribution of entanglement in a
tripartite qubit system. Since then, the monogamy relation  was
extended to other measures of quantum correlations. Unlike the
squared concurrence \cite{Coffman}, the entanglement of formation do
not satisfy the monogamy relation \cite{Coffman} in a pure
tripartite qubit system but it is satisfied in multi-mode Gaussian
state \cite{Adesso2,Adesso3}. Furthermore, quantum correlations,
measured by quantum discord, were shown to violate monogamy in some
specific quantum states \cite{Giorgi,Prabhu,Sudha, Allegra,Ren}.
Now, there are many attempts to establish the general conditions
under which a given quantum correlation measure is monogamous or not
(see \cite{Bruss} and references quoted therein). The concept  of
monogamy can be summarized as follows. Let $Q_{AB}$ denote the
shared correlation $Q$ between $A$ and $B$. Similarly, let us denote
by $Q_{A C}$ the measure of correlation between $A$ and $C$ and
$Q_{A(BC)}$ the correlation shared between $A$ and the composite
subsystem $BC$ comprising $B$ and $C$. The measure  $Q$ is
monogamous if and only if the following quantity
\begin{eqnarray}
Q_{A\vert BC} = Q_{A(BC)} -Q_{A B} - Q_{AC}
\end{eqnarray}
is positive. Therefore, quantifying the global correlation and
analyzing the monogamy of the measure $Q$ can be obtained by
quantifying  pairwise  correlations among subsystems.\\

In this work,  we derive the global quantum correlations in pure
tripartite nonorthogonal states based on the sum of correlations for
all possible bi-partitions. This is done for the widely-used
measures: concurrence, entanglement of formation, quantum discord
and geometric quantum discord. To convert the nonothogonal states to
qubits, a qubit mapping is realized. This realization is similar to
one recently used in the analysis of bipartite entanglement
properties in bipartite coherent states
\cite{Rachid1,Rachid2,Sanders,Sanders2,Sanders3,Rachid3}. As special
instance of superpositions of  nonorthogonal states, we consider
three-mode Schr\"odinger cat states, based on Glauber coherent
states. We give the explicit expressions of the global tripartite
correlations. We also discuss the limitations to sharing  quantum correlations.\\

This paper is organized as follows. In order to discuss the pairwise
quantum correlations in entangled tripartite nonorthogonal states,
we introduce, in Section 2, two different partitioning schemes. For
each scheme, a qubit mapping is proposed. In section 3, we give the
analytic expressions of pairwise entanglement of formation and
quantum discord. We discuss the conservation relation between these
two entropy based measures which implies that the tripartite measure
for quantum discord  and the entanglement of formation are
identical. In section 4, we derive the  geometric quantum discord
for all possible bipartite subsystems. As illustration, we consider
in section 5, three-mode Schr\"odinger cat states, based on Glauber
coherent states. In particular, we discuss the monogamy property of
entanglement measured by concurrence, entanglement of formation,
quantum discord and geometric quantum discord. Concluding remarks
close this paper.

\section{Tripartite nonorthogonal states}
Usually,  a tripartite state shared between three parties $A$, $B$
and $C$ is designated by a unit-trace bounded operator $\rho_{ABC}$.
In this work, we shall consider the pure tripartite state comprising
three identical subsystems living in the Hilbert space ${\cal
H}\otimes {\cal H}\otimes{\cal H}$ where ${\cal H}$ is spanned by
the set of orthonormal vectors $\{ \vert e_n \rangle : n = 1, 2,
\cdots, d\}$. The dimension  $d$ of ${\cal H}$ may be either finite
or infinite. To simplify further our purpose, we focus on tripartite
balanced entangled state of the form
\begin{equation}
\vert \Psi, m \rangle = {\cal N}(\vert \psi_1 \rangle \otimes \vert
\psi_2 \rangle  \otimes \vert \psi_3 \rangle + e^{im\pi} \vert
\phi_1 \rangle \otimes \vert \phi_2 \rangle  \otimes\vert \phi_3
\rangle ) \label{eq:main}
\end{equation}
where $m \in \bf{Z}$, $\vert \psi_i \rangle $ and $\vert \phi_i
\rangle $ are normalized states of the subsystem $i$ $(i = 1, 2,
3)$. They are linear superpositions of the eigenstates $\{ \vert e_n
\rangle\}$ of the subsystem $i$. The overlaps $\langle \psi_i \vert
\phi_i \rangle = p_i$ are in general non zero. In the equation
(\ref{eq:main}), ${\cal N}$ is given by
$$ {\cal N} = \big[ 2 + 2 p_1p_2p_3 \cos m \pi\big]^{-1/2}$$
and stands for the normalization factor of the tripartite state
$\vert \Psi, m \rangle$. We assume that  $p_1$, $p_2$ and $p_3$ are
reals. Typical examples of nonorthogonal entangled of the form
(\ref{eq:main}) are the superpositions of coherent and squeezed
states. As mentioned in the introduction, to determine the explicit
expressions of  pairwise quantum correlations present in
(\ref{eq:main}), the whole system can be partitioned in two
different ways.  For each bipartition, the bipartite states are
mapped into a two qubit systems passing from nonorthogonal states to
an orthonormal basis. This technique is similar to one used in
\cite{Fu,Wang1,Wang2,Wang3} to investigate entanglement properties
for multipartite coherent states.

\subsection{Pure bi-partitions and qubit mapping}

We first consider pure  bipartite splitting of the tripartite system
(\ref{eq:main}). In this case, the entire system splits into two
subsystems, one subsystem containing one particle and the other
containing the remaining particles. Three partitions are possible.
Indeed,  the state $\vert \Psi, m \rangle$ can be decomposed as
\begin{equation}\label{partition1}
 \vert \Psi, m
\rangle = {\cal N} (\vert \psi  \rangle_k \otimes \vert \psi
\rangle_{ij} + e^{im\pi } \vert \phi \rangle_k \otimes \vert \phi
\rangle_{ij})
\end{equation}
where
$$ \vert \psi  \rangle_k =  \vert \psi_k \rangle, \qquad  \vert \phi  \rangle_k =  \vert \phi_k \rangle
\qquad  k = 1, 2 ~{\rm or}~3,
$$
and $$\vert \psi \rangle_{ij} = \vert \psi  \rangle_i  \otimes \vert
\psi  \rangle_j \quad i,j \neq k  $$ is the state describing the
modes $i$ and $j$. The three particles state $\vert \Psi, m\rangle$
can be expressed by means of two logical qubits. This can be
realized as follows. We introduce, for the first subsystem, the
orthogonal basis $\{ \vert 0 \rangle_k , \vert 1 \rangle_k\}$
defined by
\begin{equation}\label{base1}
\vert 0 \rangle_k = \frac{ \vert \psi  \rangle_k +  \vert \phi
\rangle_k}{\sqrt{2(1 + p_k)}}
   \qquad \vert 1 \rangle_k = \frac{\vert\psi \rangle_k -  \vert \phi
\rangle_k}{{\sqrt{2(1-  p_k)}}}.
\end{equation}
Similarly, we introduce, for the second subsystem $(ij)$, the
orthogonal basis $\{ \vert  0 \rangle_{ij} , \vert  1
\rangle_{ij}\}$ given by
\begin{equation}\label{base2}
\vert 0 \rangle_{ij} = \frac{ \vert \psi \rangle_{ij} +  \vert \phi
\rangle_{ij}}{\sqrt{2(1 +p_ip_j)}}
   \qquad \vert 1 \rangle_{ij} = \frac{\vert \psi \rangle_{ij} -  \vert \phi
\rangle_{ij}}{{\sqrt{2(1- p_ip_j)}}}.
\end{equation}
Inserting(\ref{base1}) and (\ref{base2}) in (\ref{partition1}), we
get the form of the pure state $ \vert \Psi, m \rangle$ in the basis
$\{ \vert 0 \rangle_{k} \otimes \vert 0 \rangle_{ij} ,
 \vert 0 \rangle_{k} \otimes \vert 1 \rangle_{ij} , \vert  1 \rangle_{k}
 \otimes \vert 0 \rangle_{ij} , \vert  1 \rangle_{k} \otimes \vert  1
 \rangle_{ij}\}$. Explicitly, it is given by
\begin{equation}
 \vert \Psi, m
\rangle = \sum_{\alpha= 0,1} \sum_{\beta= 0,1} C_{\alpha,\beta}
\vert \alpha \rangle_k \otimes \vert \beta
\rangle_{ij}\label{mapping1}
\end{equation}
where the coefficients $C_{\alpha,\beta}$ are
$$ C_{0,0} = {\cal N}(1 + e^{im\pi}) c^+_{k}c^+_{ij}  , \qquad  C_{0,1} =  {\cal N} (1 -e^{im\pi}) a^+_{k}c^-_{ij} $$
$$ C_{1,0} = {\cal N} (1 - e^{im\pi}) c^+_{ij}c^-_{k}  , \qquad  C_{1,1} =  {\cal N} (1 + e^{im\pi}) c^-_{k}c^-_{ij}. $$
in terms of the quantities
$$c^{\pm}_k =\sqrt{\frac{1 \pm p_k}{2}} \qquad  c^{\pm}_{ij} =\sqrt{\frac{1 \pm p_ip_j}{2}} $$
involving the scalar products $p_i$ between the nonorthogonal states
$\vert \psi_i \rangle$ and  $\vert \phi_i \rangle$.

\subsection{Mixed bi-partitions and qubit mapping}

The second partition can be realized by considering  the bipartite
reduced density matrix $\rho_{ij}$ which is obtained by tracing out
the degrees of freedom of the third subsystem $k$:
\begin{equation}\label{partition2}
\rho_{ij} = \text{Tr}_{_{{\rm k\neq i,j}}}(\vert \Psi, m \rangle
\langle \Psi, m \vert).
\end{equation}
In this case, three different bipartite mixed states are also
possible: $\rho_{12}$ , $\rho_{13}$ and $\rho_{23}$. The reduced
density matrix  $\rho_{ij}$ is given by
\begin{eqnarray}
\rho_{ij} &=&{\cal N}^2(\vert \psi_i , \psi_j \rangle \langle \psi_i
, \psi_j \vert +\vert \phi_i , \phi_j \rangle \langle \phi_i ,
\phi_j | + e^{i m \pi } q_{ij} |\phi_i , \phi_j\rangle \langle
\psi_i , \psi_j \vert +e^{-i m \pi }q_{ij} \vert \psi_i , \psi_j
\rangle \langle \phi_i, \phi_j \vert ) \label{rhoij}
\end{eqnarray}
with $q_{ij} \equiv p_1p_2p_3/p_ip_j$. It is interesting to note
that the density $\rho_{ij}$ is a  rank-2 mixed state. Indeed, the
state (\ref{rhoij}) can be written  as
\begin{eqnarray}
\rho_{ij} = \frac{{\cal N}^2} {{\cal N}_{ij}^2}~ \bigg[a_{ij}^2~
\vert \Psi_{ij} \rangle \langle \Psi_{ij} \vert  +  b_{ij}^2  ~Z
\vert \Psi_{ij} \rangle \langle \Psi_{ij} \vert Z\bigg]
\label{rhoijsecondrank}
\end{eqnarray}
where ${\cal N}_{ij}$ is the normalization factor of the bipartite
state $\vert \Psi_{ij} \rangle $ given by
$$ \vert \Psi_{ij} \rangle = {\cal N}_{ij} (\vert \psi_i , \psi_j \rangle + e^{i m \pi} \vert \phi_i , \phi_j \rangle)$$
and the operator $Z$ is the third Pauli generator defined by
$$ Z \vert \Psi_{ij} \rangle = {\cal N}_{ij} (\vert \psi_i , \psi_j \rangle - e^{i m \pi} \vert \phi_i , \phi_j \rangle).$$
The coefficients $a_{ij}$ and $b_{ij}$ occurring in
(\ref{rhoijsecondrank}) are expressed in terms of the quantities
$q_{ij}$ as follows
$$ a_{ij} = \sqrt{\frac{1 + q_{ij}}{2}} \qquad b_{ij} = \sqrt{\frac{1 - q_{ij}}{2}}.$$
Here also, one can map the reduced system $\rho_{ij}$ into a pair of
two-qubits. As hereinabove, we define, for the subsystem $i$, the
orthogonal basis $\{\vert {\bf 0}_i\rangle ,\vert {\bf 1}_i\rangle
\}$ by
\begin{equation}
\vert\psi_i\rangle \equiv  a_i \vert {\bf 0}_i \rangle + b_i \vert
{\bf 1}_i \rangle \qquad \vert \phi_i\rangle\equiv a_i \vert {\bf
0}_i \rangle - b_i
 \vert {\bf 1}_i \rangle~,\label{basei}
\end{equation}
where $$a_i = \sqrt{\frac{1+p_i}{2}} \qquad b_i =
\sqrt{\frac{1-p_i}{2}}.$$ Similarly, we introduce, for the subsystem
$j$, a second two dimensional orthogonal basis as
\begin{equation}
\vert\psi_j\rangle \equiv  a_j \vert {\bf 0}_j \rangle + b_j \vert
{\bf 1}_j \rangle \qquad \vert \phi_j\rangle\equiv a_j \vert {\bf
0}_j \rangle - b_j
 \vert {\bf 1}_j \rangle~,\label{basej}
\end{equation}
where $$a_j = \sqrt{\frac{1+p_j}{2}} \qquad b_j =
\sqrt{\frac{1-p_j}{2}}.$$ Substituting Eqs.~(\ref{basei}) and
(\ref{basej}) into Eq.~(\ref{rhoij}), it is simple to reexpress the
2-rank mixed density (\ref{rhoijsecondrank}) in the two qubit basis
$\{\vert{\bf 0}_i{\bf 0}_j\rangle ,\vert{\bf 0}_i{\bf 1}_j\rangle
,\vert{\bf 1}_i{\bf 0}_j\rangle ,
    \vert{\bf 1}_i{\bf 1}_j\rangle \}$.
The  pure as well as mixed bi-partitions and the qubit mappings
introduced in this section provides us with a simple way to derive
the pairwise quantum correlations and subsequently the global
quantum correlations in multipartite nonorthogonal states. This is
discussed in the following sections.

\section{Quantum discord and entanglement of formation in tripartite nonorthogonal states}

\subsection{Bipartite measures of entanglement of formation and quantum discord}
The total correlation  in a quantum state $\rho_{AB}$ is quantified
by the mutual information
\begin{equation}\label{def: mutual information}
    I_{AB}=S_{A}+S_{B}-S_{AB},
\end{equation}
where $\rho_{AB}$ is the state of a bipartite quantum system
composed of the subsystems $A$ and $B$, the operator
$\rho_{A(B)}={\rm Tr}_{B(A)}( \rho_{AB})$ is the reduced state of
$A$($B$) and $S(\rho)$ is the von Neumann entropy of a quantum state
$\rho$. The mutual information $I_{AB}$ contains both quantum and
classical correlations. It can be decomposed as
$$ I_{AB} =  D_{AB} +  C_{AB}.$$
Consequently, for a bipartite quantum system, the quantum discord
$D_{AB}$ is defined as the difference between total correlation
$I_{AB}$ and classical correlation $C_{AB}$. The classical part
$C_{AB}$ can be determined by a local measurement optimization
procedure as follows. Let us consider a perfect measurement on the
subsystem $A$ defined by a positive operator valued measure (POVM).
The set of POVM elements is denoted by $\mathcal{M}=\{M_k\}$ with
$M_k\geqslant 0$ and $\sum_k M_k= \mathbb{I} $. The von Neumann
measurement,  on the subsystem $A$, yields the statistical ensemble
$\{ p_{B,k} , \rho_{B,k}\}$ such that
$$\rho_{AB} \longrightarrow \frac{(M_k \otimes \mathbb{I})\rho_{AB}(M_k \otimes \mathbb{I})}{p_{B,k}}$$
where the measurement operation is written as \cite{Luo}
\begin{eqnarray}
M_k = U \, \Pi_k \, U^\dagger \label{Eq:VNmsur}
\end{eqnarray}
with $\Pi_k = |k\rangle\langle k| ~ (k = 0,1)$ is the one
dimensional projector for subsystem $A$  along the computational
basis $|k\rangle$,  $U \in SU(2)$ is a unitary operator and
$$ p_{B,k} = {\rm Tr}  \bigg[ (M_k \otimes \mathbb{I})\rho_{AB}(M_k \otimes \mathbb{I}) \bigg]. $$
The amount of information acquired about particle $B$ is then given
by
$$S_{B}-\sum_k ~p_{B,k} ~S_{B,k},$$
which depends on measurements belonging to $\mathcal{M}$. To remove
the measurement dependence, a maximization over all possible
measurements is performed and the classical correlation writes
\begin{eqnarray}
    C_{AB}& =\max_{\mathcal{M}}
    \Big[S_{B}-\sum_k ~p_{B,k} ~S_{B,k}\Big] \nonumber \\
    & =S(\rho_B) - \widetilde{S}_{\rm min}
      \label{def: classical correlation}
\end{eqnarray}
where $\widetilde{S}_{\rm min}$  denotes the minimal value of the
conditional  entropy
\begin{equation}\label{condit-entropy}
\widetilde{S} =  \sum_k ~p_{B,k} ~S_{B,k}.
\end{equation}
When optimization is taken over all perfect measurement, the quantum
discord is
\begin{equation} \label{def: discord}
    D_{AB} \equiv D^{\rightarrow}_{AB} = I_{AB} - C_{AB}
    =S_{A}+\widetilde{S}_{\rm min}-S_{AB}.
\end{equation}
Thus,  the derivation of quantum discord requires the minimization
of conditional entropy. This constitutes a complicated issue when
dealing with an arbitrary mixed state. The explicit analytical
expressions of quantum discord were obtained only for few
exceptional two-qubit quantum states, especially ones of rank two.
One may quote for instance the results obtained in
\cite{Ali,Adesso2} (see also \cite{Rachid1,Rachid2,Rachid3}). For a
density matrix of rank two, the minimization of the conditional
entropy (\ref{condit-entropy}) can be performed by purifying the
density matrix $\rho_{AB}$ and  making use of Koashi-Winter relation
\cite{Koachi-Winter} (see also \cite{Shi1}). This relation
establishes the connection between the classical correlation of a
bipartite state $\rho_{AB} $ and the entanglement of formation of
its complement $\rho_{BC}$.  Hereafter, we discuss briefly this nice
relation. For a  rank-two quantum state, the density matrix
$\rho_{AB}$ decomposes as
\begin{eqnarray}
\rho_{AB} = \lambda_+ \vert \phi_+ \rangle \langle \phi_+ \vert +
\lambda_- \vert \phi_- \rangle \langle \phi_- \vert
\end{eqnarray}
where  $\lambda_+$ and $\lambda_-$ are the eignevalues of $\rho_{AB}
$ and   the corresponding eigenstates are denoted by $\vert \phi_+
\rangle$ and $\vert \phi_- \rangle$ respectively. Attaching a qubit
$C$ to the two-qubit system $A$ and $B$, the purification of the
system yields
\begin{eqnarray}
\vert \phi \rangle = \sqrt{\lambda_+} \vert \phi_+ \rangle \otimes
\vert {\bf 0}  \rangle +  \sqrt{\lambda_-} \vert \phi_- \rangle
\otimes \vert {\bf 1} \rangle
\end{eqnarray}
such that the whole system $ABC$ is described by the pure state
$\rho_{ABC} = \vert \phi \rangle \langle \phi \vert $ from which one
has  the bipartite densities $\rho_{AB} = {\rm Tr}_C \rho_{ABC}$ and
$\rho_{BC} = {\rm Tr}_A \rho_{ABC}$. According to Koachi-Winter
relation \cite{Koachi-Winter}, the minimal value of the conditional
entropy coincides with the entanglement of formation of $\rho_{BC}$.
It is given by
\begin{equation}\label{stild-min}
 \widetilde{S}_{\rm min} = E(\rho_{BC}) = H(\frac{1}{2} + \frac{1}{2} \sqrt{1 - \vert {\cal C}(\rho_{BC})\vert^2})
\end{equation}
where  $H(x) = -x\log_{2} x -(1-x)\log_{2} (1-x)$ is the binary
entropy function and ${\cal C}(\rho_{BC})$ is the concurrence of the
density $\rho_{BC}$. We recall that for $\rho_{12}$ the density
matrix for a pair of qubits~$1$ and~$2$ which may be pure or mixed,
the concurrence is~\cite{Wootters98}
\begin{equation}
{\cal C}_{12}=\max \left\{ \lambda _1-\lambda _2-\lambda _3-\lambda
_4,0\right\} \label{eq:c1}
\end{equation}
for~$\lambda_1\ge\lambda_2\ge\lambda_3\ge\lambda_4$ the square roots
of the eigenvalues of the "spin-flipped" density matrix
\begin{equation}
\varrho_{12}\equiv\rho_{12}(\sigma_y\otimes\sigma
_y)\rho_{12}^{\star}(\sigma_y\otimes \sigma_y), \label{eq:c2}
\end{equation}
where the star stands for complex conjugation in the basis $\{ \vert
00 \rangle, \vert 01 \rangle, \vert 10 \rangle, \vert 11 \rangle \}$
and $\sigma_y $ is the usual Pauli matrix. It  follows that the
Koaschi-Winter relation and the purification procedure provide us
with a computable expression of quantum discord
\begin{equation}
D^{\rightarrow}_{AB} =  S_A - S_{AB} + E_{BC}
\end{equation}
when the measurement is performed on the subsystem $A$. In the same
manner, performing measurement on the second subsystem $B$, one gets
\begin{equation}
D^{\leftarrow}_{AB} = S_B - S_{AB} + E_{AC}.
\end{equation}
It is simple to check that for a pure density state $\rho_{AB}$, the
quantum discord reduces to entanglement of formation given by the
entropy of the reduced density of the subsystem $A$.

\subsection{Quantum discord in pure tripartite nonorthogonal states }
In the pure bi-partitioning  scheme (\ref{partition1}), using the
Wootters concurrence formula (\ref{eq:c1}), it is simply verified
that
\begin{equation} \label{conc-pure}
 {\cal C}_{k (ij)} =
\frac{\sqrt{(1 - p_k^2)(1- p_i^2p_j^2)}}{ 1 + p_1p_2p_3 \cos m \pi}.
\end{equation}
It follows that the entanglement of formation  writes
\begin{equation}\label{E-pure}
 E_{k (ij)} =  H\bigg(\frac{1}{2} + \frac{1}{2} ~\frac{ p_k + p_ip_j\cos m\pi }{1 + p_1p_2p_3\cos m\pi }\bigg)
\end{equation}
and coincides with the quantum discord
\begin{equation}\label{D-pure}
 E_{k (ij)} =  D_{k (ij)}.
\end{equation}
For the mixed states $\rho_{ij}$ associated with the second
partitioning scheme (\ref{rhoij}), the concurrence (\ref{eq:c1})
takes the following form
\begin{equation}\label{conc-mixte}
{\cal C}_{ij}= q_{ij}~\frac{\sqrt{(1 - p_i^2)(1- p_j^2)}}{ 1 +
p_1p_2p_3 \cos m \pi}.
\end{equation}
The pairwise quantum discord present in the mixed states $\rho_{ij}$
can be computed using the procedure presented in the previous
subsection. As result,  when  the measurement is performed on the
subsystem $A \equiv i$,  the quantum discord is
\begin{equation}\label{disc-mixte}
 D^{\rightarrow}_{ij} =  S_i - S_{ij} + E_{jk}
\end{equation}
where $k$ stands for the third subsystem traced out to get the
reduced matrix density $\rho_{ij}$. The von Neumann entropy of the
reduced density $\rho_i$ is
\begin{equation}\label{Si}
S_i = H\bigg(\frac{1}{2} \frac{(1 + p_i)(1 + p_jq_{ij}\cos m\pi)}{1
+p_1p_2p_3\cos m\pi}\bigg),
\end{equation}
and the entropy of the bipartite density $\rho_{ij}$ is explicitly
given by
\begin{equation}\label{Sij}
S_{ij} = H\bigg(\frac{1}{2} \frac{(1 + p_ip_j\cos m\pi)(1 +
q_{ij})}{1 +p_1p_2p_3\cos m\pi}\bigg).
\end{equation}
It important to emphasize that the entanglement of formation
measuring the entanglement of the subsystem $j$ with the ancillary
qubit, required in the purification process to minimize the
conditional entropy, is exactly the entanglement of formation
measuring the degree of intricacy between the subsystem $j$ and the
traced out qubit $k$. It is given by
\begin{equation}\label{Ejk}
E_{jk}  = H\bigg(\frac{1}{2} + \frac{1}{2} \sqrt{ 1- \frac{p_i^2(1 -
p_j^2)(1 - p_{k}^2)}{(1 +p_1p_2p_3\cos m\pi)^2}}\bigg).
\end{equation}
Using the equations (\ref{Si}), (\ref{Sij}) and (\ref{Ejk}), one
obtains
\begin{equation}\label{Dij}
 D^{\rightarrow}_{ij} = H\bigg(\frac{(1 + p_i)(1 + p_jq_{ij}\cos m\pi)}{2(1
+p_1p_2p_3\cos m\pi)}\bigg)- H\bigg( \frac{(1 + p_ip_j)(1 +
q_{ij}\cos m\pi)}{2(1 +p_1p_2p_3\cos m\pi)}\bigg)+
H\bigg(\frac{1}{2} + \frac{1}{2} \sqrt{ 1- \frac{p_i^2(1 - p_j^2)(1
- q_{ij}^2)}{(1 +p_1p_2p_3\cos m\pi)^2}}\bigg) ,
\end{equation}
Also, because the whole system is pure, we have
\begin{equation}\label{Sij=Sk}
S_{ij} = S_k \qquad i,j \neq k.
\end{equation}
Using the equations
(\ref{Si}), (\ref{Sij}) and (\ref{Ejk}), one obtains the following
conservation relation
$$D^{\rightarrow}_{12} + D^{\rightarrow}_{23}+  D^{\rightarrow}_{31} = E_{12} + E_{13} + E_{23},$$
reflecting that the sum of the bipartite quantum discord present in
all mixed states $\rho_{ij}$ is exactly the sum of the bipartite
entanglement of formation. It is important to notice  that the
conservation law for the distribution of entanglement of formation
and quantum discord, in a pure tripartite system,  was firstly
derived in \cite{Fanchini2,Fanchini}. Similarly, the explicit form
of the quantum, when performing a measurement on the qubit $j$, is
$$ D^{\leftarrow}_{ij} = S_j - S_{ij} + E_{ik}$$
and we have the following asymmetric relation
\begin{equation}\label{asym-rel}
 D^{\leftarrow}_{ij} = D^{\rightarrow}_{ji}.
\end{equation}
The quantum discord $ D^{\leftarrow}_{ij}$ (resp.
$D^{\rightarrow}_{ij}$) is the portion of the mutual information in
the the bipartite state $\rho_{ij}$ that is locally inaccessible by
$i$ (resp. $j$). In this sense quantum discord can be interpreted as
the fraction of the pairwise mutual information which can not be
accessible by a local measurement. Based on the asymmetry definition
of quantum discord, two useful quantities can introduced:
\cite{Fanchini}
$$\Delta^+_{ij} = \frac{1}{2} \big( D^{\rightarrow}_{ij} + D^{\leftarrow}_{ij}\big)\qquad \Delta^-_{ij} = \frac{1}{2} \big( D^{\rightarrow}_{ij} - D^{\leftarrow}_{ij}\big).$$
The sum $\Delta^+_{ij}$ is the average of locally inaccessible
information when the measurements are performed on the subsystems
$i$ and $j$. It quantifies the disturbance caused by any local
measurement. The difference $\Delta^-_{ij}$ was termed by Fanchini
et al \cite{Fanchini} the balance of locally inaccessible
information and quantifies the asymmetry between the subsystems in
responding to the measurement disturbance. Using the expressions of
quantum discord given by (\ref{Dij}) and the asymmetric relation
(\ref{asym-rel}), one  verifies that the quantities $\Delta^+_{ij}$
and $\Delta^-_{ij}$ satisfy the following distribution relations
\begin{equation} \label{delta+}
\Delta^+_{12} + \Delta^+_{13} + \Delta^+_{23} = E_{12} + E_{13} +
E_{23},
\end{equation}
and
\begin{equation}\label{delta-}
\Delta^-_{12} + \Delta^-_{13} + \Delta^-_{23} = 0.
\end{equation}
Consequently, using the results (\ref{D-pure}) and (\ref{delta+}),
the global quantum correlation (\ref{Qtotal}) when bipartite
correlations are measured by quantum discord writes
\begin{equation}
D_{(1,2,3)} = \frac{1}{6} \bigg(E_{12} + E_{13} + E_{23} + E_{1(23)}
+ E_{2(13)} + E_{3(12)}\bigg).
\end{equation}
This shows that the sum of quantum discord for all possible
partitions coincides the global entanglement of formation
\begin{equation}
D_{(1,2,3)} = E_{(1,2,3)}.
\end{equation}

\section{ Geometric  quantum discord in tripartite nonorthogonal state}
\subsection{ Definition}
The geometric measure of quantum discord  is defined as the distance
between a state $\rho$ of a bipartite system $AB$ and the closest
classical-quantum state presenting zero discord \cite{Dakic2010}:
\begin{equation}\label{dg}
  D^{g}(\rho):=\min_{\chi}||\rho-\chi||^{2}
\end{equation}
where the minimum is over the set of zero-discord states $\chi$ and
the distance is the square norm in the Hilbert-Schmidt space. It is
given by
$$||\rho-\chi||^{2}:= {\rm Tr}(\rho-\chi)^2. $$
When the measurement is taken on the subsystem $A$, the zero-discord
state $\chi$ is represented as \cite{Ollivier-PRL88-2001}
$$\chi= \sum_{i = 1,2}p_{i}|\psi_{i}\rangle\langle\psi_{i}|\otimes\rho_{i}
$$
where $p_i$ is a probability distribution, $\rho_{i}$ is the
marginal density matrix of $B$ and  $\{|\psi_1\rangle
,|\psi_2\rangle \}$ is an arbitrary orthonormal vector set. An
arbitrary two qubit state writes in Bloch representation as
\begin{eqnarray}
  \rho & = & \frac{1}{4}\left[ \sigma_{0}\otimes \sigma_{0} +\sum_{i}^{3}(x_{i}\sigma_{i}\otimes \sigma_{0}
   +y_{i} \sigma_{0}\otimes\sigma_{i})+\sum_{i,j=1}^{3}R_{ij}\sigma_{i}\otimes\sigma_{j}\right]
\end{eqnarray}
where $x_{i} = {\rm Tr}\rho(\sigma_{i}\otimes \sigma_{0}),~  y_{i} =
{\rm Tr}\rho(\sigma_{0}\otimes\sigma_{i})$ are the components of
local Bloch vectors and $R_{ij} = {\rm
Tr}\rho(\sigma_{i}\otimes\sigma_{j})$ are components of the
correlation tensor. The operators $\sigma_i$ $( i = 1, 2, 3)$ stand
for the three Pauli matrices and $\sigma_0$ is the identity matrix.
The explicit expression of the geometric  quantum discord is given
by \cite{Dakic2010}:
\begin{equation}
  D^g(\rho)=\frac{1}{4}\left(||x||^{2}+||R||^{2}-k_{\rm {max}}\right) \label{eq:GMQD_original}
\end{equation}
where $x=(x_{1},x_{2},x_{3})^{T}$, $R$ is the matrix with elements
$R_{ij}$ and $k_{\rm{max}}$ is the largest eigenvalue of the matrix
defined by
\begin{equation}
K := xx^{T}+RR^{T}. \label{matrix K}
\end{equation}
Denoting the eigenvalues of the $3\times 3$ matrix $K$ by
$\lambda_1$, $\lambda_2$ and $\lambda_3$ and considering
$||x||^{2}+||R||^{2}={\rm Tr}K$, we get an alternative compact form
of the geometric measure of quantum discord \cite{Rachid3}
\begin{equation}
D^g(\rho) = \frac{1}{4}~ {\rm min}\{ \lambda_1 + \lambda_2 ,
\lambda_1 + \lambda_3 , \lambda_2 + \lambda_3\} \label{eq:GMQD_new}
\end{equation}
which is more convenient for our purpose.

\subsection{Geometric measure of quantum discord for the pure bipartite states}

Using the tools presented in the previous subsection, we shall
determine the global geometric quantum discord in the tripartite
state (\ref{eq:main}). We evualuate first the pairwise geometric
discord in the  pure bipartite states  (\ref{partition1}). For this,
using the Schmidt decomposition decomposition, we write the state
$\vert \Psi, m \rangle$ as
\begin{equation}
\vert \Psi, m \rangle = \sqrt{\lambda_+} ~\vert + \rangle_k \otimes
\vert + \rangle_{ij} + \sqrt{\lambda_-}~ \vert - \rangle_k \otimes
\vert - \rangle_{ij} \label{shcmidt}
\end{equation}
where $\vert \pm \rangle_k$ denotes the eigenvectors of the reduced
density matrix  associated with the first subsystem containing the
particle $k$. Similarly, $\vert \pm \rangle_{ij}$ denotes the
eigenvectors of the reduced density matrix  for the second subsystem
comprising the particles  $i$ and $j$.  The eigenvalues
$\lambda_{\pm}$ are given by
$$ \lambda_{\pm} = \frac{1}{2} \bigg(1 \pm \sqrt {1 - {\cal C}_{k(ij)}^2}\bigg)$$
where the bipartite concurrence ${\cal C}_{k(ij)}$ is given by the
equation (\ref{conc-pure}). In this case,  the matrix $K$, defined
by (\ref{matrix K}), takes the diagonal form
$$ K = {\rm diag} (4 \lambda_+ \lambda_-, 4 \lambda_+ \lambda_-, 2 ( \lambda_+^2 + \lambda_-^2) ),$$
and using  the equation (\ref{eq:GMQD_new}), the pairwise geometric
discord is given by
\begin{equation}
D^g_{k(ij)}=\frac{1}{2} \frac{(1- p_k^2)(1-p_{i}^2p_{j}^2)}{(1+
p_1p_2 p_3\cos m\pi)^2}\label{discordpure}
\end{equation}
It is remarkable that the geometric quantum discord can be
re-expressed as
\begin{equation}
D^g_{k(ij)}= \frac{1}{2}~{\cal C}_{k(ij)}^2\label{qdpurestate}
\end{equation}
in terms of  the bipartite concurrence ${\cal C}_{k(ij)}$. This
equation traduces the relation between the geometric discord and the
concurrence for pure bipartite states.
\subsection{Geometric measure of quantum discord for mixed bipartite states}
Having derived the geometric discord in the pure bipartition scheme,
we now consider the mixed states of the form (\ref{rhoij}) obtained
in the second bipartition scheme. In this order, we write the matrix
$\rho_{ij}$ as follows
\begin{equation}
\rho_{ij} = \sum_{\alpha \beta} R_{\alpha \beta}
\sigma_{\alpha}\otimes \sigma_{\beta}
\end{equation}
where the non vanishing correlation matrix elements $R_{\alpha
\beta}$ $(\alpha, \beta = 0,1,2,3)$ are given by
$$ R_{00} = 1, \quad R_{11} = 2{\cal N}^2 \sqrt{(1- p_i^2)(1- p_j^2)}, \quad R_{22} = -2{\cal N}^2 \sqrt{(1- p_i^2)(1- p_j^2)}~p_{k}\cos
m\pi,$$ $$ R_{33} = 2{\cal N}^2 (p_ip_j + p_{k}\cos m\pi), \quad
R_{03} = 2{\cal N}^2 (p_j + p_i p_{k}\cos m\pi), \quad R_{30} =
2{\cal N}^2 (p_i + p_j p_{k}\cos m\pi).$$ In this case, the
eigenvalues of the matrix $K$ (\ref{matrix K}) write
\begin{equation}
\lambda_1 = 4 {\cal N}^4 \bigg[(1 + p_i^2)(p_j^2 + p_{k}^2) + 4
(p_1p_2p_3) \cos m\pi\bigg]\label{lambda1}
\end{equation}
\begin{equation}
\lambda_2 = 4 {\cal N}^4 (1 - p_i^2)(1 - p_j^2)\label{lambda2}
\end{equation}
\begin{equation}
\lambda_3 = 4 {\cal N}^4 (1 - p_i^2)(1 -
p_j^2)p_{k}^2\label{lambda3}
\end{equation}
Noticing that   $0 \leq p_i \leq 1$, it is easy to see that
$\lambda_3 \leq \lambda_2$. Thus, the equation (\ref{eq:GMQD_new})
reduces to
\begin{equation}
D^g_{ij} = \frac{1}{4} {\rm min}\{  \lambda_1 + \lambda_3 ,
\lambda_2 + \lambda_3\}.\label{Dgplus}
\end{equation}
Subsequently, for  the mixed states $\rho_{ij}$, the explicit
expression of geometric quantum discord writes
\begin{equation}
D^g_{ij} =  \frac{1}{4} \frac{(1 - p_i^2)(1 - p_j^2)(1+p_{k}^2)}{(1
+ p_1p_2 p_3\cos m\pi)^2}\label{Dgplus-general}
\end{equation}
 when the condition
$\lambda_1 > \lambda_2$ is satisfied or
\begin{equation}
D^g_{ij} = \frac{1}{4}\frac{(1 + p_i^2)(p_j^2 + p_{k}^2)+ (1 -
p_i^2)(1 - p_j^2)p_{k}^2 + 4 (p_1p_2 p_3) \cos m\pi}{(1 + p_1p_2
p_3\cos m\pi)^2} \label{Dgmoins-general}
\end{equation}
in the  situation where $ \lambda_1 < \lambda_2$.\\
Finally the measure of multipartite  quantum correlation
(\ref{Qtotal}) for geometric quantum discord, in the pure tripartite
state (\ref{eq:main}), writes
\begin{equation}
D^g_{(1,2,3)} = \frac{1}{6} \bigg(D^g_{12} + D^g_{21} +D^g_{13} +
D^g_{31} +D^g_{23}+D^g_{23} \bigg) + \frac{1}{12} \bigg({\cal
C}^2_{1(23)} + {\cal C}^2_{2(13)} + {\cal C}^2_{3(12)}\bigg).
\end{equation}

\section{ Illustration: three-mode Schr\"odinger cat states}
To illustrate the results obtained in the previous sections, we need
to consider a specific instance of tripartite system involving non
orthogonal states. In this sense, we consider a three-mode
Schr\"odinger cat state
\begin{equation}\label{cat-states}
 \vert \alpha, m  \rangle = {\cal N}_m(\vert \alpha \vert)\bigg( \vert \alpha \rangle_1 \vert \alpha \rangle_2 \vert \alpha
 \rangle_3
+ e^{im\pi} \vert -\alpha \rangle_1 \vert -\alpha \rangle_2 \vert
-\alpha \rangle_3\bigg),
\end{equation}
based on Glauber  or radiation field coherent states $\vert \alpha
\rangle$
\begin{equation}
|\alpha\rangle = e^{-\frac{|\alpha|^2}{2}} \sum_{n=0}^{\infty}
\frac{\alpha^n}{\sqrt{n!}}|n\rangle
\end{equation}
where the complex number $\alpha$ characterizes the amplitude of the
coherent state $\vert \alpha \rangle$ and $\vert n \rangle$ is a
Fock state (also known as a number state). The normalization factor
in (\ref{cat-states}) is given by
$${\cal N}_m(\vert \alpha \vert) = ( 2 + 2 e^{-6|\alpha|^2} \cos
m\pi)^{-\frac{1}{2}}.$$ Considering this special tripartite state
involving Glauber coherent states, we shall in what follows give the
global quantum correlations $Q_{(1,2,3)}$ (see eq.(\ref{Qtotal}))
when the pairwise correlations are measured by the squared
concurrence, entanglement of formation, entropy based quantum
discord or its geometrized variant. Furthermore, this specific
tripartite state allows us to decide about the monogamy of each of
these measures.

Two interesting limits of the Schr\"odinger cat states
(\ref{cat-states}) arise when $\alpha \rightarrow \infty$ and $
\alpha \rightarrow 0$. We first consider the asymptotic limit
$\alpha \rightarrow \infty$. In this limit  the two states $|\alpha
\rangle $ and $|- \alpha \rangle $ approach orthogonality, and an
orthogonal basis can be constructed such that $\vert {\bf
0}\rangle\equiv \vert \alpha \rangle$ and $\vert{\bf 1}\rangle
\equiv \vert - \alpha \rangle$. Thus, the state $ \vert \alpha , m
\rangle$ approaches a multipartite state of ${\rm GHZ}$ type
\begin{equation}
\vert \alpha , m\rangle \sim \vert {\rm GHZ}\rangle_3 = \frac
1{\sqrt{2}}(\vert {\bf 0}\rangle \otimes |{\bf 0}\rangle
\otimes\vert {\bf 0}\rangle
    +e^{i m \pi}\vert {\bf 1}\rangle \otimes
    \vert {\bf 1}\rangle \otimes
\vert {\bf 1}\rangle).\label{GHZ}
\end{equation}
In the situation where $\alpha \rightarrow 0$, one should
distinguish separately the cases $m = 0 ~({\rm mod}~2)$ and $m = 1
~({\rm mod}~2)$. For $m$ even, the tripartite superposition
(\ref{cat-states}) reduces to ground state
\begin{equation}
\vert 0 , 0 ~({\rm mod}~ 2)  \rangle \sim  \vert 0\rangle
\otimes\vert 0 \rangle  \otimes \vert 0 \rangle,
\end{equation}
and  for $m$ odd, the state $\vert \alpha , 1 ~({\rm mod}~ 2)
\rangle$ reduces to  a multipartite state of W type~\cite{Dur00}
\begin{equation}
\vert 0 ,  1 ~({\rm mod}~ 2) \rangle \sim \vert\text{\rm W}\rangle_3
    = \frac{1}{\sqrt{3}}(\vert 1\rangle \otimes\vert 0 \rangle \otimes
       \vert0\rangle  +\vert 0\rangle \otimes\vert 1\rangle \otimes\vert0\rangle
+ \vert 0\rangle \otimes\vert 0\rangle  \otimes  \vert 1\rangle)~.
\label{Wstate}
\end{equation}
Here $\vert n\rangle $ $(n=0,1)$ denote the Fock-Hilbert states.

It follows that the states $\vert \alpha, m = 0 ~({\rm mod}~2),
\rangle$ interpolate between states of ${\rm GHZ}$ type $(\alpha
\rightarrow \infty)$ and the separable state $\vert 0\rangle
\otimes\vert 0 \rangle \otimes \vert 0 \rangle$ $(\alpha \rightarrow
0)$. In other hand, the states $\vert \alpha , m = 1 ~({\rm mod}~2),
\rangle$ may be viewed as interpolating between states of ${\rm
GHZ}$ type $(\alpha \rightarrow \infty)$ and states of W type
$(\alpha \rightarrow 0)$.


\subsection{ Global quantum correlations and monogamy relation}

\subsubsection{ Squared concurrence}

Using the equation (\ref{conc-pure}) and noticing that the states
$\rho_{1(23)}$,$\rho_{2(13)}$  and $\rho_{3(12)}$ are identical, it
is simple to check that the concurrences in the pure bipartite
splitting are all equals. Explicitly, they are given by
\begin{equation}\label{C123-pure}
{\cal C}_{1(23)} = {\cal C}_{2(13)}= {\cal C}_{3(12)}
=\frac{\sqrt{(1 - p^2)(1- p^4)}}{ 1 + p^3 \cos m \pi}.
\end{equation}
where $p = \langle \alpha \vert -\alpha \rangle = e^{-2\vert \alpha
\vert^2}$. In the the second bipartite splitting (\ref{partition2}),
the mixed density matrices $\rho_{12}$,$\rho_{23}$  and $\rho_{13}$
are identical and  the concurrence (\ref{conc-mixte}) rewrites
\begin{equation}\label{C12-mixte}
{\cal C}_{12} = {\cal C}_{23}= {\cal C}_{13} =\frac{p(1 - p^2)}{ 1 +
p^3 \cos m \pi}.
\end{equation}
To examine the monogamy relation of entanglement measured by the
concurrence in quantum systems involving three qubits, Coffman et al
\cite{Coffman} introduced the so called three tangle defined as
follows
\begin{equation}\label{tangle}
\tau_{i\vert jk} =  {\cal C}_{i(jk)}^2 - {\cal C}_{ij}^2 - {\cal
C}_{ik}^2.
\end{equation}
Reporting (\ref{C123-pure}) and (\ref{C12-mixte}) in (\ref{tangle}),
one gets
$$
\tau_{1\vert 23} = \tau_{2\vert 13} = \tau_{3\vert 12} \equiv \tau
$$
with
$$\tau = \frac{(1 - p^2)^2(1- p)^2}{ (1 + p^3 \cos m \pi)^2}. $$
The three tangle $\tau $   is always positive. This result reflects
the monogamy of entanglement measured by the squared concurrence. In
other hand, using the expressions (\ref{C123-pure}) and
(\ref{C12-mixte}) and replacing the pairwise quantum correlation $Q$
in (\ref{Qtotal}) by the squared concurrence, the global tripartite
quantum correlation (\ref{Qtotal})  in the tripartite Schr\"odinger
cat states (\ref{cat-states}) takes the following form
$${\cal C}^2_{(1,2,3)} =  \frac{1}{2}  \frac{(1+2p^2)(1-p^2)^2}{(1 + p^3\cos m\pi)^2}.$$

\subsubsection{ Entanglement of formation and quantum
discord}

As above, to decide about the monogamy of entanglement measured by
the entanglement of formation, we introduce the  following quantity
\begin{equation} \label{E-tangle}
E_{i\vert jk} =  E_{i(jk)} - E_{ij} - E_{ik}.
\end{equation}
For the Schr\"odinger cat states under consideration, the pairwise
entanglement of formation corresponding to the pure bipartition
scheme (\ref{partition1}) can be obtained from equation
(\ref{E-pure}). One gets
\begin{equation} \label{E123-cat}
E_{1(23)} = E_{2(13)}= E_{3(12)} = H\bigg(\frac{1}{2} + \frac{1}{2}
~\frac{ p + p^2\cos m\pi }{1 + p^3\cos m\pi }\bigg)
\end{equation}
In the second splitting scheme (\ref{partition2}), we have
$\rho_{12}=\rho_{23}=\rho_{13}$. In this case, the equation
(\ref{Ejk}) gives
\begin{equation}\label{Eij-cat}
E_{12} = E_{23}= E_{13} =  H\bigg(\frac{1}{2} + \frac{1}{2} \sqrt{
1- \frac{p^2(1 - p^2)^2}{(1 + p^3\cos m\pi)^2}}\bigg).
\end{equation}
Substituting the expressions (\ref{E123-cat}) and (\ref{Eij-cat}) in
the equation (\ref{E-tangle}), one obtains
$$E_{1\vert 23} = E_{ 2 \vert 13}  = E_{ 3\vert 12} \equiv E $$
where the quantity $E$ is given by
$$E =  H\bigg(\frac{1}{2} + \frac{1}{2}
~\frac{ p + p^2\cos m\pi }{1 + p^3\cos m\pi }\bigg) - 2
H\bigg(\frac{1}{2} + \frac{1}{2} \sqrt{ 1- \frac{p^2(1 - p^2)^2}{(1
+ p^3\cos m\pi)^2}}\bigg).$$ The behavior of the quantity $E$ vs the
overlap $p$ is depicted in the  following figure.
\begin{center}
\includegraphics[width=3in]{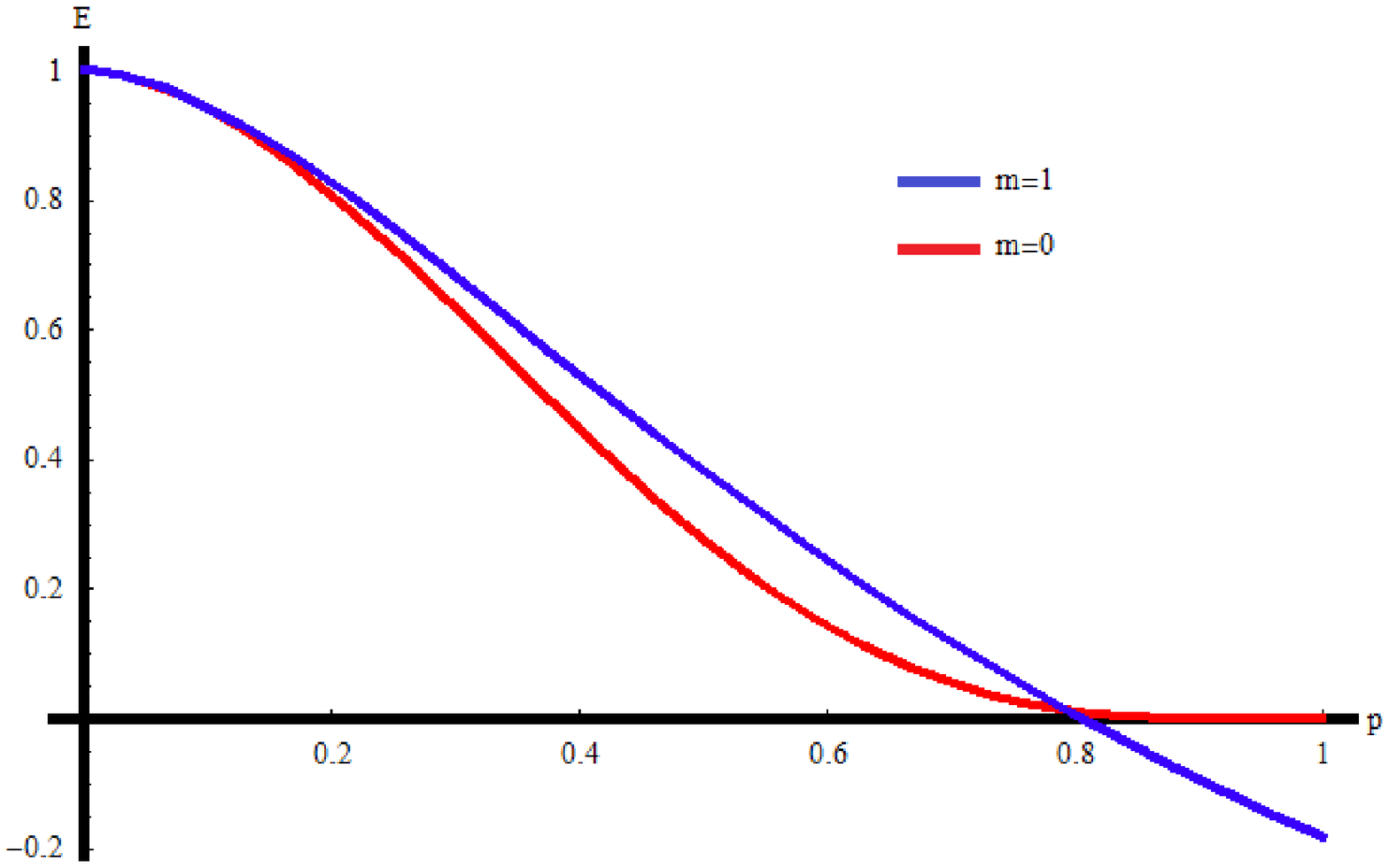}\\
{\bf Figure 1}  {\sf $E = E_{i\vert jk}$ versus the overlapping $p$
 for  $m=0$ and $m=1$.}
\end{center}
Clearly, the entanglement of formation is monogamous for symmetric
three modes Schr\"odinger cat states $(m = 0)$ for any value of $p$.
The antisymmetric states $(m = 1)$  possess monogamy property only
when $0 \leq p \lesssim 0.8$. The figure 3 reveals that the $\vert
GHZ \rangle_3$ state ($p \rightarrow 0$)
follows monogamy and $\vert W \rangle_3$ state ($p \rightarrow 1$)  does not.\\
The sum of the pairwise entanglement of formation,  in all possible
bi-partitions, is then given by
\begin{equation}\label{E123}
E_{(1,2,3)} =    \frac{1}{2}  \bigg[ H\bigg(\frac{1}{2} +
\frac{1}{2} \sqrt{ 1- \frac{p^2(1 - p^2)^2}{(1 + p^3\cos
m\pi)^2}}\bigg) + H\bigg(\frac{1}{2} + \frac{1}{2} ~\frac{ p +
p^2\cos m\pi }{1 + p^3\cos m\pi }\bigg) \bigg].
\end{equation}

To compute the global amount of pairwise quantum discord in the
states (\ref{cat-states}) and to investigate the monogamy relation ,
two important remarks are in order. First, note that in a pure state
the entanglement of formation and quantum discord coincide. In this
respect, in the pure bipartition scheme (\ref{partition1}), one has
$$E_{1\vert 23} = D_{1\vert 23}  \quad  E_{ 2 \vert 13}  =D_{ 2 \vert 13} \quad E_{ 3\vert 12} =  D_{ 3\vert 12}$$
Furthermore, using the equations (\ref{Ejk}) and (\ref{Dij}), one
can verify that for the reduced mixed states
$\rho_{12}=\rho_{13}=\rho_{23}$, the entanglement of entanglement of
formation coincides with quantum discord. Indeed, we have
$$E_{12} = D_{12} \quad E_{23}=  D_{23} \quad E_{13} = D_{13}. $$
It is remarkable that the bipartite mixed states $\rho_{12}$
$\rho_{13}$ and  $\rho_{23}$ constitute a special class of mixed
states where entanglement of formation coincides with quantum
discord. Thus, the measures of entanlement of formation and  quantum
discord, in the Schr\"odinger cat states (\ref{cat-states}), are
identical and the global amount of quantum discord coincides, as
expected, with the global entanglement of formation  given by
(\ref{E123}).

\subsubsection{ Geometric quantum discord}

Now, we consider the global quantum correlation measured by
geometric quantum discord. For the states (\ref{cat-states}), from
the equation (\ref{discordpure}), one has
$$
D^g_{1(23)}= D^g_{2(13)}= D^g_{3(12)}
$$
with
\begin{equation}\label{Dg123}
D^g_{1(23)}= \frac{1}{2}~{\cal C}_{1(23)}^2 = \frac{1}{2}~\frac{(1 -
p^2)(1 - p^4)}{(1 + p^3\cos m\pi)^2}.
\end{equation}
For the mixed states $\rho_{12}$, $\rho_{13}$ and $\rho_{23}$ which
are identical, we treat the symmetric and anti-symmetric cases
separately. For $ m = 0$, using (\ref{Dgplus-general}), the
geometric quantum discord writes
\begin{equation}\label{Dg12+}
D^g_{12}= D^g_{23}= D^g_{13}=
\frac{1}{4}~\frac{p^2(1+p)^2(2+(1-p)^2)}{(1+p^3)^2}
\end{equation}
for $0 \leq p \leq \sqrt{2} - 1$ and from (\ref{Dgmoins-general})
one obtains
\begin{equation}\label{Dg12-}
D^g_{12}= D^g_{23}= D^g_{13}= \frac{1}{4}~\frac{(1+
p^2)(1+p)^2(1-p)^2}{(1+p^3)^2}
\end{equation}
when $\sqrt{2} - 1 \leq p \leq 1 $. For the antisymmetric
Schr\"odinger cat states $( m = 1)$, the geometric quantum discord
is
\begin{equation}
D^g_{12}= D^g_{23}= D^g_{13}=
\frac{1}{4}~\frac{p^2(2+(1+p)^2)}{(1+p+p^2)^2}
\end{equation}

It follows that, for even tripartite Schr\"odinger cat states $(m =
0)$, the total amount of quantum correlation measured by the
geometric discord is
$$ D^g_{(1,2,3)} =  \frac{1}{8} \frac{(1+p)^2(2p^2 + (1-p^2)(2+3p^2))}{(1+p^3)^2}$$
for $ 0 \leq p \leq  \sqrt{2} - 1,$ and
$$ D^g_{(1,2,3)} =  \frac{3}{8} \frac{(1+p^2)(1-p^2)^2}{(1+p^3)^2}$$
when  $\sqrt{2} - 1 \leq p \leq  1 $. For odd Schr\"odinger cat
states $(m = 1)$, the sum of all possible pairwise  geometric
quantum discord is given by the following equation
$$ D^g_{(1,2,3)}) =  \frac{1}{8} \frac{2p^2 + (1+p)^2(2+3p^2)}{(1+p+p^2)^2}$$
for $ 0 \leq p \leq   1.$\\
Note that the maximal value of geometric discord (\ref{dg}) for two
qubit states is 1/2 and it is not normalized to one. Hence, for
comparison with the others normalized measures, we consider $2D^g$
as a proper measure.

In the figures 2 and 3, a comparison of tripartite quantum
correlation for the squared concurrence, usual quantum discord and
its geometrized version are represented. Figure 2 displays that
these three measures give approximatively the same amount of quantum
correlation for $m = 0$. This corroborates the fact that the
entanglement of formation, quantum discord and geometric quantum
discord possess the monogamy property like the squared concurrence.
Figure 3 reveals that for $m = 1$, the sum of entanglement of
formation (or equivalently the usual quantum discord) becomes larger
than the sum of pairwise quantum correlations measured by the
concurrence and the geometric discord, especially when $p$
approaches the unity. Furthermore, the global sum of squared
concurrences behaves like the sum of bipartite geometric discord for
$ 0 \leq p \leq 0.5$ and increases slowly after but the behavior
stays slightly the same as geometric discord.

\begin{center}
\includegraphics[width=3in]{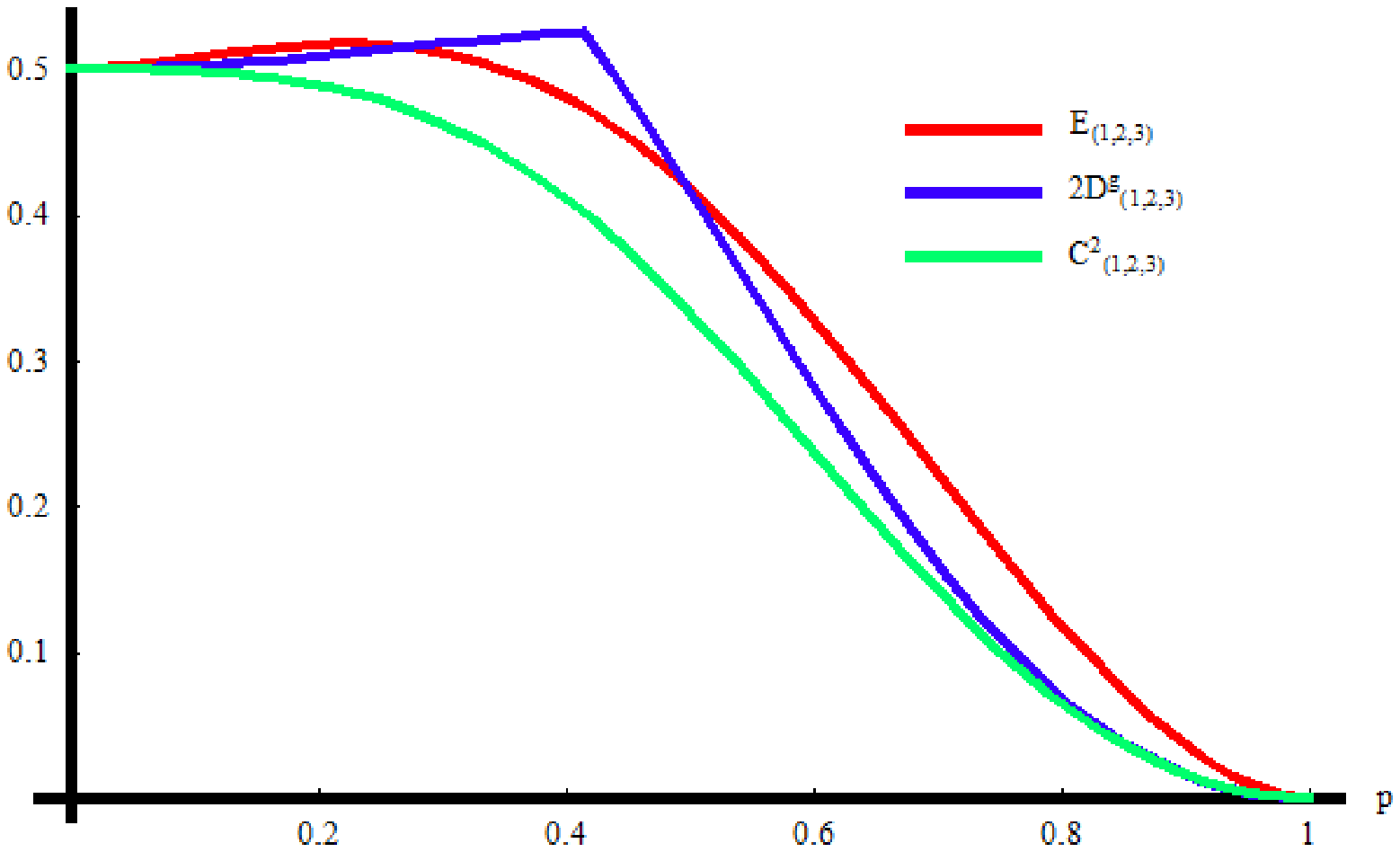}\\
{\bf Figure 2}  {\sf  Tripartite quantum correlation versus the
overlapping $p$ for  $m=0$.}
\end{center}

\begin{center}
\includegraphics[width=3in]{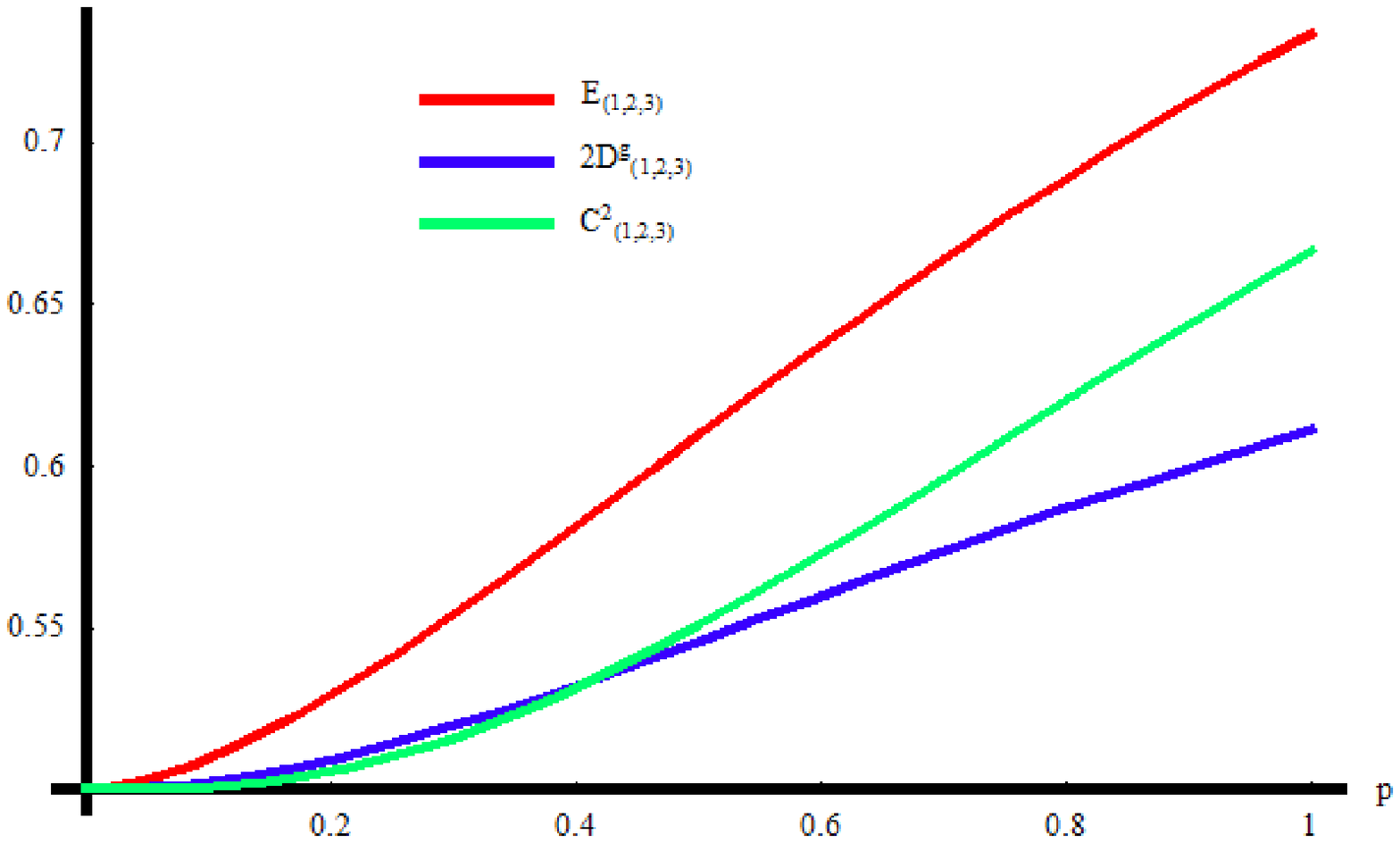}\\
{\bf Figure 3} {\sf  Tripartite quantum correlation versus the
overlapping $p$ for   $m=1$.}
\end{center}

Finally, to examine the monogamy of geometric quantum discord, one
should analyzes the positivity of the  following quantity
$$ D^g_{i\vert jk}  =  D^g_{i(jk)} - D^g_{ij} - D^g_{ik}. $$
For the tripartite cat states (\ref{cat-states}), we have
$$ D^g_{1\vert 23} =  D^g_{2\vert 13} =  D^g_{3\vert 12} \equiv D^g . $$
In the symmetric case $(m = 0)$, the quantity $D^g$ vanishes for
$\sqrt{2}-1 \leq p \leq 1$ and it is given by
$$D^g = \frac{1}{2}~ \frac{(1+p)^2(1 - (\sqrt{2}+1)p)(1 - (\sqrt{2}-1)p)}{(1 + p^3)^2}$$
for $ 0 \leq p \leq \sqrt{2}-1 $. It is simple to verify that in
this case the geometric discord is monogamous. For antisymmetric
Schr\"odinger cat states $(m = 1)$, one obtains
$$D^g = \frac{1}{2} ~\frac{(1+ 2p - p^2)}{(1 + p +  p^2)^2},$$
which is always positive.  In this respect, The geometric quantum
discord follows the monogamy property for any value of the overlap
$p$.


\section{ Concluding remarks}

In summary, we have explicitly derived the quantum correlation in a
tripartite system involving nonorthogonal states. The total amount
of quantum correlation is defined  as the sum of all pairwise
quantum correlations. It is evaluated using measures which go beyond
entanglement, e.g., usual quantum discord and its geometrized
version. A suitable qubit mapping was realized for all possible
bi-partitions of the system. We have shown that the sum of all
pairwise entanglement of formation in a pure entangled tripartite
state is exactly the sum of pairwise quantum discord of all possible
bi-partitions. This peculiar result originates from the conservation
relation between the entanglement of formation and quantum discord.
We also examined the monogamy relation of concurrence, entanglement
of formation, quantum discord and quantum discord in the special
case of  non orthogonal three-modes Schr\"odinger cat states. We
proved that squared concurrence and geometric discord are
monogamous. The entanglement of formation and quantum discord
follows  the monogamy property in the symmetric tripartite
Schr\"odinger cat states $(m=0)$. However, in the antisymmetric case
($m=1$), they cease to be monogamous when the three-mode cat states
approache the three qubit states $W_3$ corresponding to the
situation where $p \rightarrow 1$. The odd Schr\"odinger cat states
(\ref{cat-states}) interpolate continuously between the ${\rm GHZ}$
type states (\ref{GHZ}) ($p \rightarrow 0$) and ${\rm W}$ states
(\ref{Wstate}) ($p \rightarrow 1$). The ${\rm GHZ}$ states maximize
the pure entanglement of formation $E_1(23)$ between any qubit and
the two others. The ${\rm W}$ states maximize the entanglement of
formation $E_{12}$ in the mixed states obtained after tracing out
the third
qubit.\\

Finally, It must be noticed that the investigation of monogamy and
polygamy of quantum correlations in multipartite quantum systems is
deeply dependent on the choice of correlations measures. Many
exciting issues, regarding this problem, remain open. The
quantification of the genuine multipartite correlations constitutes
a key challenge in the field of quantum information theory to
understand the distribution of correlations in quantum systems
comprising many parts.


\begin{thebibliography}{99}

\bibitem{NC-QIQC-2000} M.A. Nielsen and I.L. Chuang, {\it Quantum Computation and Quantum Information} (Cambridge Univ. Press, Cambridge,
2000).

\bibitem{Alber-QI2001} G. Alber, T. Beth, M. Horodecki, P. Horodecki, R. Horodecki, M. R\"otteler, H. Weinfurter, R. Werner and A. Zeilinger,
 {\it Quantum Information} (Springer-Verlag, Berlin, 2001), ch. 5.

\bibitem{Vedral-RMP-2002} V. Vedral, Rev. Mod. Phys. {\bf 74} (2002)
197.

\bibitem{Horodecki-RMP-2009} R. Horodecki, P. Horodecki, M. Horodecki and K. Horodecki, Rev. Mod. Phys. {\bf 81}(2009) 865.

\bibitem{Guhne} O. G\"uhne and G. T\'oth, Phys. Rep. {\bf 474} (2009) 1.

\bibitem{vedral-modi} K. Modi, A. Brodutch, H. Cable, T. Paterek and V. Vedral, Rev. Mod. Phys. {\bf 84} (2012) 1655.

\bibitem{Ben1} C.H. Bennett, G. Brassard, C. Cr\'epeau, R. Jozsa, A. Peres and W.
K. Wootters,  Phys. Rev. Lett. {\bf 70} (1993) 1895.

\bibitem{Ben2} C.H. Bennett, G. Brassard, Proceedings of the IEEE International
Conference on Computers, Systems, and Signal Processing (IEEE, New
York, 1984, Bangalore, India,, 1984), pp 175-179.

\bibitem{Eckert} A.K. Ekert, Phys. Rev. Lett. {\bf 67} (1991) 661.

\bibitem{Murao} M. Murao, D. Jonathan, M.B. Plenio and V. Vedral, Phys. Rev. A
{\bf 59} (1999) 156.


\bibitem{Ollivier-PRL88-2001} H. Ollivier and W.H. Zurek, Phys. Rev. Lett. {\bf 88} (2001) 017901.

\bibitem{Vedral-et-al} L. Henderson and V. Vedral, J. Phys. A {\bf 34}(2001)  6899;
V. Vedral, Phys. Rev. Lett. {\bf 90}  (2003) 050401; J. Maziero, L.
C. Cel\'eri, R.M. Serra and V. Vedral, Phys. Rev A {\bf 80} (2009)
044102.


\bibitem{Luo} S. Luo, Phys. Rev. A \textbf{77} (2008) 042303;  Phys.
Rev. A \textbf{77} (2008) 022301.

\bibitem{Ali} M. Ali, A.R.P. Rau and G. Alber,  Phys. Rev. A {\bf 81} (2010) 042105.

\bibitem{Shi1} M. Shi, W. Yang, F. Jiang and J. Du, J. Phys. A: Mathematical and Theoretical
{\bf 44} (2011) 415304.


\bibitem{Girolami} D. Girolami and G. Adesso, Phys. Rev. A {\bf 83} (2011)
052108.

\bibitem{Shi2} M. Shi, F. Jiang, C. Sun and J. Du, New Journal of Physics {\bf 13} (2011)
073016.

\bibitem{Rachid1}  M. Daoud and R. Ahl Laamara, J. Phys. A: Math. Theor. {\bf 45} (2012) 325302.

\bibitem{Rachid2}  M. Daoud and R. Ahl Laamara,
International Journal of Quantum Information {\bf 10} (2012)
1250060.

\bibitem{Dakic2010} B. Dakic, V. Vedral and C. Brukner, phys. Rev.
Lett. {\bf 105} (2010) 190502.

\bibitem{Wootters98} W.K. Wootters, Phys. Rev. Lett. {\bf 80} (1998) 2245; W.K. Wootters,  Quant. Inf. Comp. {\bf 1} (2001)
27.

\bibitem{Zhou} D.L. Zhou, B. Zeng, Z. Xu, and L. You, Phys. Rev. A {\bf 74} (2006)
052110.

\bibitem{Kaszlikowski} D. Kaszlikowski,  A. Sen(De), U. Sen, V. Vedral, and A. Winter,
Phys. Rev. Lett. {\bf 101} (2008) 070502.

\bibitem{Bennett} C.H. Bennett, A. Grudka, M. Horodecki, P. Horodecki, and R.
Horodecki, Phys. Rev. A {\bf 83} (2011) 012312.

\bibitem{Giorgi1} G.L. Giorgi, B. Bellomo, F. Galve, and R.
Zambrini, Phys. Rev. Lett. {\bf 107} (2011) 190501.

\bibitem{Li-Luo} N. Li and S. Luo, Phys. Rev. A {\bf 84} (2011)
042124.

\bibitem{Hald} J. Hald, J.L. Sørensen, C. Schori and E.S. Polzik, Phys. Rev.
Lett. {\bf 83} (1999) 1319.

\bibitem{Kuzmich} A. Kuzmich, L. Mandel and N.P. Bigelow, Phys. Rev. Lett. {\bf 85}
(2000) 1594.

\bibitem{Meyer} V. Meyer, M. A. Rowe, D. Kielpinski, C.A. Sackett, W.M. Itano, C.
Monroe and D.J. Wineland, Phys. Rev. Lett. {\bf 86} (2001) 5870.


\bibitem{Chakrabarty} I. Chakrabarty, P. Agrawal and A.K. Pati, The European Physical
Journal D  {\bf 65} (2011) 605.

\bibitem{Rulli} C.C. Rulli and M.S. Sarandy, Phys. Rev. A {\bf 84} (2011)
042109.

\bibitem{Z-H Ma} Z-H Ma, Z-H Chen and F.F. Fanchini, New Journal of Physics, {\bf 15}
(2013) 043023.

\bibitem{Coffman}  V. Coffman, J. Kundu and W.K. Wootters, Phys. Rev.
A {\bf 61} (2000) 052306.

\bibitem{Adesso2} G. Adesso and F. Illuminati, New J. Phys. {\bf 8} (2006)
15.

\bibitem{Adesso3} T.Hiroshima, G. Adesso, and F. Illuminati, Phys. Rev.
Lett. {\bf 98} (2007) 050503.

\bibitem{Giorgi} G.L. Giorgi, Phys. Rev. A {\bf 84} (2011) 054301.

\bibitem{Prabhu}  R. Prabhu, A.K. Pati, A.S. De and U. Sen, Phys. Rev. A
{\bf 86} (2012) 052337.

\bibitem{Sudha} Sudha, A.R. Usha Devi and A.K. Rajagopal, Phys. Rev. A {\bf 85}
(2012) 012103.

\bibitem{Allegra}  M. Allegra, P. Giorda   and A. Montorsi, Phys. Rev. B {\bf 84}
(2011) 245133.

\bibitem{Ren} X.-J. Ren and H. Fan, Quant. Inf. Comp. Vol. {\bf 13} (2013) 0469.

\bibitem{Bruss} A. Streltsov, G. Adesso, M. Piani and D. Bruss, Phys. Rev.
Lett. {\bf 109} (2012) 050503.

\bibitem{Sanders} B.C. Sanders, Phys. Rev. A {\bf 45} (1992) 6811.

\bibitem{Sanders2}  B.C. Sanders, Phys. Rev. A {\bf 46} (1992) 2966.

\bibitem{Sanders3} B.C Sanders, J. Phys. A: Math. Theor. {\bf 45} (2012) 244002.

\bibitem{Rachid3} M. Daoud and R. Ahl Laamara,
 Phys. Lett. A {\bf 376} (2012) 2361.

\bibitem{Fu} H. Fu, , X. Wang and  A.I. Solomon, Phys. Lett. A {\bf 291}(2001)
73.

\bibitem{Wang1} X. Wang, J. Phys. A: Math. Gen. {\bf 35} (2002) 165.

\bibitem{Wang2} X. Wang and B.C. Sanders, Phys. Rev. A {\bf 68} (2003) 012101.

\bibitem{Wang3} X. Wang and B.C. Sanders, Phys. Rev. A {\bf 65} (2002) 012303.

\bibitem{Koachi-Winter} M. Koachi and A. Winter, Phys. Rev. A {\bf
69} (2004) 022309.


\bibitem{Fanchini2} F.F. Fanchini, M.F. Cornelio, M.C. de Oliveira and A.O. Caldeira,
Phys. Rev. A {\bf 84} (2011) 012313.

\bibitem{Fanchini} F.F. Fanchini, L.K. Castelano, M.F. Cornelio, M.C. de Oliveira,  New Journal of Physics {\bf 14} (2012)
013027.

\bibitem{Dur00} W. D\"ur, G. Vidal and J.I. Cirac, Phys. Rev. A
{\bf 62} (2000) 062314.






\end{thebibliography}
\end{document}